\documentclass[aps,pra,scriptaddress,twocolumn,noshowkeys,noshowpacs,superscriptaddress]{revtex4-1}
\usepackage[T1]{fontenc}
\usepackage[latin1]{inputenc}
\usepackage{lmodern}
\expandafter\let\csname equation*\endcsname\relax
\expandafter\let\csname endequation*\endcsname\relax
\usepackage{amsmath}
\usepackage{amssymb}
\usepackage{graphicx}
\usepackage{xcolor}
\usepackage{natbib}
\usepackage{bm}
\usepackage{bbold}
\usepackage[mathscr]{euscript}
\usepackage[cal=boondoxo]{mathalfa}
\usepackage{comment}
\renewcommand\footnotemark{}

\newcommand{\g}{\mbox{\scriptsize 1D}}

\newcommand{\inter}{\mbox{\scriptsize int}}
\newcommand{\ind}{\mbox{\scriptsize ind}}
\newcommand{\eff}{\mbox{\scriptsize eff}}

\begin{document}
\title{Population mixing due to dipole-dipole interactions in a 1D array of multilevel atoms}

\author{E. Munro}
\affiliation{Centre for Quantum Technologies, National University of Singapore, 3 Science Drive 2, 117543 Singapore.}
\affiliation{Entropica Labs, 32 Carpenter Street, 059911 Singapore.}
\author{A. Asenjo-Garcia}
\affiliation{Norman Bridge Laboratory of Physics MC12-33, California Institute of Technology, Pasadena, CA 91125, USA.}
\affiliation{Institute for Quantum Information and Matter, California Institute of Technology, Pasadena, CA 91125, USA.}
\affiliation{ICFO-Institut de Ciencies Fotoniques, The Barcelona Institute of Science and Technology, 08860 Castelldefels (Barcelona), Spain.}
\author{Y. Lin}
\affiliation{JILA, National Institute of Standards and Technology and University of Colorado, Boulder, Colorado 80309, USA.}
\author{L. C. Kwek}
\affiliation{Centre for Quantum Technologies, National University of Singapore, 3 Science Drive 2, 117543 Singapore.}
\affiliation{Institute of Advanced Studies, Nanyang Technological University, 60 Nanyang View, Singapore 639673.}
\affiliation{National Institute of Education, Nanyang Technological University, 1 Nanyang Walk, Singapore 637616.}
\affiliation{MajuLab, CNRS-UNS-NUS-NTU International Joint Research Unit, UMI 3654, Singapore.}
\author{C. A. Regal}
\affiliation{JILA, National Institute of Standards and Technology and University of Colorado, Boulder, Colorado 80309, USA.}
\affiliation{Department of Physics, University of Colorado, Boulder, Colorado 80309, USA.}
\author{D. E. Chang}
\affiliation{ICFO-Institut de Ciencies Fotoniques, The Barcelona Institute of Science and Technology, 08860 Castelldefels (Barcelona), Spain.}
\affiliation{ICREA-Instituci\'{o} Catalana de Recerca i Estudis Avan\c{c}ats, 08015 Barcelona, Spain.}

\date{\today}
\begin{abstract}
We examine theoretically how dipole-dipole interactions arising from multiple photon scattering lead to a modified distribution of ground state populations in a driven, ordered one-dimensional array of multilevel atoms. Specifically, we devise a level configuration in which a ground-state population accumulated solely due to dipole-dipole interactions can be up to 20\% in regimes accessible to current experiments with neutral atom arrays. For much larger systems, the steady state can consist of an equal distribution of population across the ground state manifold. Our results illustrate how dipole-dipole interactions can be accentuated through interference, and regulated by the geometry of ordered atom arrays. More generally, control techniques for multilevel atoms that can be degraded by multiple scattering, such as optical pumping, will benefit from an improved understanding and control of dipole-dipole interactions available in ordered arrays.
\end{abstract}
\maketitle

\section{INTRODUCTION}

Atomic ensembles are a prevailing platform for quantum light-matter interfaces, with applications in quantum information processing, metrology, and nonlinear optics \cite{HammererReview}. Conventionally, the interaction of an atomic ensemble with light is modeled by the semi-phenomenological Maxwell-Bloch equations \cite{Bonifaccio_MBEqns,McCallHahn_SIT}, where the propagation of an electromagnetic field mode of interest is described by a quasi-one-dimensional (1D) wave equation, while the coupling of the atoms to all other modes is assumed to yield independent spontaneous emission. On the other hand, interference and multiple scattering in dense three-dimensional (3D) ensembles can produce highly fundamental yet complex phenomena, of which a complete theoretical understanding remains to be developed. Examples include the linear optical response and refractive index of dense ensembles \cite{fleischhauer1999radiative,BrowaeysOpticalResonanceShifts,CoherentScatteringNearResonantLightBrowaeys,schilder2017homogenization,corman2017transmission}, Anderson localization \cite{AndLoc_Germans,AndLoc_Russians}, radiation trapping \cite{RadTrapping1986Paper,KaiserRadiationTrapping,KaiserRadTrppingMultilevel}, superradiance \cite{GrossHarocheSuprad,ScullySuprad2008,Superrad_Strathclyde} and subradiance \cite{KaiserSubradiance}.

Separately, experimental advances now allow highly ordered arrays of $N\sim 100$ neutral atoms to be assembled in 1D and 2D, providing a novel platform for controlled investigation of fundamental atom-light interactions in microscopic, atom-by-atom detail \cite{lester2015rapid,EndresAtomAssembly,ahn,Browaeys2D,Lukin51Qubit,browaeysarxiv}. The inherent periodicity in such systems can produce strong interference in the emitted fields, and highly non-trivial optical phenomena can occur, including subradiance in the form of guided modes in 1D chains \cite{RitschSubadiance2017,SelectiveSubradiance}, strong reflection of incident fields from 2D arrays \cite{Bettles2DReflection,Shahmoon2DReflection}, and topological edge states \cite{PerczelTopological,bettles2017topological}.

\begin{figure}
\centering
\includegraphics[width=0.33\textwidth]{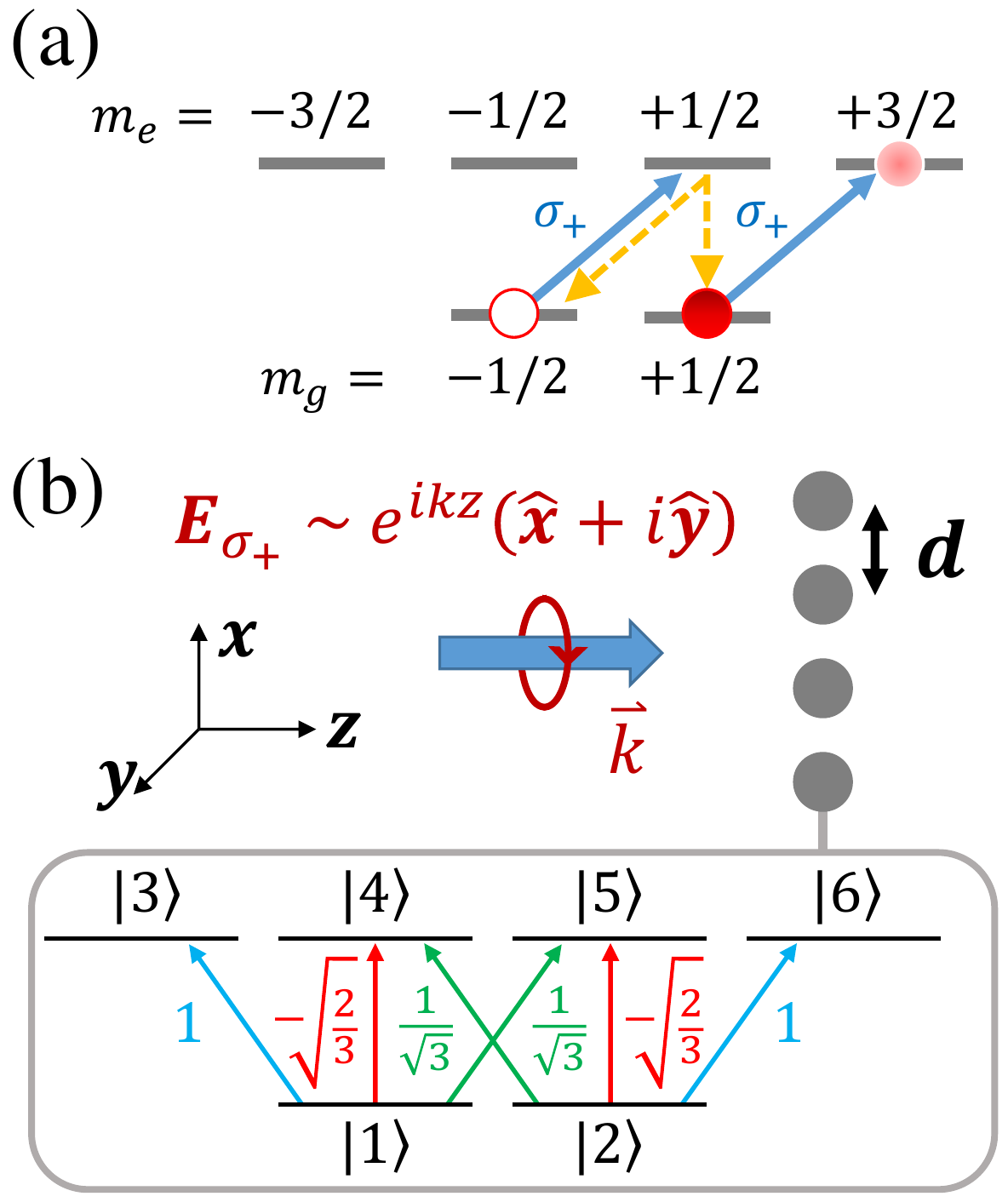}
\caption{(a) An atom with an initial distribution of ground state populations is driven by a circularly polarized incident field (blue arrows). Population in the $m_g = -1/2$ state is raised to the $m_e = +1/2$ state, following which it returns to the $m_g = -1/2$ state, or decays to the $m_g = +1/2$ state (dashed yellow arrows). In the long-time limit, all population is pumped into the subspace containing the $m_g = +1/2$ and $m_e = +3/2$ `stretched' states (solid circles), while the $m_g = -1/2$ state is unoccupied (empty circle).
(b) Schematic illustration of our system. An atomic chain is oriented along the $x$ axis, with lattice constant $d$. The atomic internal structure is shown in the box, with states labeled $|1\rangle$ through $|6\rangle$. Colored arrows (numbers) depict the electric dipole-allowed transitions (corresponding Clebsch-Gordan coefficients). The atoms are driven with a $\sigma_+$-polarized field (electric field rotating clockwise with respect to the positive $z$ axis) of wavevector $\vec{k} = k\mathbf{\hat{z}}$ propagating in the $z$ direction, which coincides with the atomic quantization axis.}
\label{SystemSetup}
\end{figure}

While these previous analyses considered atoms with only a single ground state, in this Letter we show how dipole-dipole interactions (DDIs) can manifest themselves in the steady-state population distribution of a driven 1D array of multilevel atoms. Specifically, we consider atoms with a manifold of ground and excited states of total angular momenta $J_g=1/2$ and $J_e = 3/2$ respectively, whose Zeeman sublevels are labeled by the magnetic quantum numbers $m_g$ and $m_e$. When a single such atom is driven with a circularly polarized field, the atomic population is pumped entirely into the two-level subspace spanned by the `stretched' states with $m_g = +1/2$ and $m_e = +3/2$ (the states of maximum angular momentum in each manifold) -- see Fig. \ref{SystemSetup}(a). In particular, the steady-state population of the $m_g = -1/2$ sublevel is identically zero.

While this independent-atom description may apply to dilute or spatially unordered atomic ensembles, in an ordered 1D array multiple photon scattering can strongly modify the distribution of atomic populations, even for lattice spacings $d$ exceeding a resonant wavelength $\lambda$. Specifically, we show that in arrays containing a modest number of atoms $N$, and where $d = m\lambda$ for integer $m$, the steady-state population of the $m_g=-1/2$ sublevel is $\sim(\lambda/d)^2\log^2 N$, and can be of order 10 - 20\% under conditions accessible to current experiments. Meanwhile, as $N\rightarrow\infty$ the ratio of populations in the two ground states approaches unity.

Our results are appealing because they illustrate how quantum optical phenomena arising from strong DDIs can be first explored in neutral atom array experiments. Specifically, our scheme does not require sub-wavelength lattice spacings, as would be needed to observe subradiance in 1D arrays ($d<\lambda/2$, \cite{RitschSubadiance2017,SelectiveSubradiance}) or full reflection from 2D arrays ($d<\lambda$, \cite{GarciaDeAbajoReview,Bettles2DReflection,Shahmoon2DReflection}). Moreover, our proposal explicitly accounts for -- and relies on -- multilevel ground-state structure, a situation relevant to existing experimental setups.

Furthermore, by illuminating the relationship between multiple scattering and ground state populations, our results provide insight into the efficiency of techniques such as optical pumping \cite{kastler1950,HapperOPReview} and laser cooling \cite{Phillips1998Review} in dense ensembles \cite{WiemanReabsorption1991,Kerman3DMolasses} and ordered arrays. Optical pumping -- widely used for initializing atoms in well-defined ground sublevels -- is essential for protocols in quantum computation, information processing, and precision measurement, where fiducial state preparation is required \cite{DiVincenzo,BlattIonTrapQIP,JunYeOpticalClocksReview}. Our techniques also provide a way to scale between small ($N \sim 2$) systems, where the effect of multiple scattering on populations can be studied exactly \cite{KiffnerGeometryLambda,KiffnerBreakdown}, to large systems \cite{Happer1987}, where it is qualitatively known that such effects can be relevant, but have been historically difficult to quantitatively model.

\section{MODELLING OF SYSTEM}

We consider a 1D chain of multilevel atoms extended along the $x$ direction, perpendicular to the atomic quantization axis ($z$ axis). The ground and excited manifolds support two ($m_g=\pm 1/2$) and four ($m_e = \pm 1/2, \pm 3/2$) Zeeman sublevels [Fig. \ref{SystemSetup}(a)]. To lighten the notation, we denote these sublevels in order as $g = \{1,2\}$ and $e = \{3,4,5,6\}$ [Fig. \ref{SystemSetup}(b)]. The atoms are driven by a $\sigma_+$-polarized incident field propagating along the $z$ axis. While we focus on this specific configuration, we emphasize that our theoretical model may be applied to any arrangement of atoms, in any dimension, with any internal structure. However, the system considered here is elegant in its simplicity, while strong collective effects are expected to be prominent under conditions accessible to current experiments, as we now explain qualitatively.

As noted above, absent multiple scattering the steady-state population for our configuration is entirely within the stretched state subspace. In particular, the population of state $|1\rangle$ is identically zero. However, photon reabsorption can lead to population being acquired in all magnetic sublevels. To appreciate this, first note that the input field, whose polarization vector is $\hat{\mathbf{\epsilon}}_L = \hat{\mathbf{\sigma}}_+ =  -(\hat{\mathbf{x}} + i\hat{\mathbf{y}})/\sqrt{2}$, induces an atomic dipole with both $x$ and $y$ components. Along the array axis, the $x$ component produces no radiated field, and the far field is purely $y$-polarized. Since this constitutes a superposition of circularly polarized components $\hat{\sigma}_{\pm}=\mp (\hat{\mathbf{x}}\pm i\hat{\mathbf{y}})/\sqrt{2}$, the atoms are also driven by a field component that instead pumps to state $|1\rangle$, giving a distribution of population across all levels. Intuitively, this effect should be enhanced where the scattered fields interfere constructively, such as when the lattice spacing is an integer multiple of the resonant wavelength $\lambda$.

We emphasize that our driving configuration and system geometry are chosen to accentuate the effect of DDIs. Specifically, the atomic dipoles will remain active under continuous application of the driving field. In contrast, in a typical optical pumping configuration the atoms would be driven into a dark ground state with respect to the incoming field, and the DDIs would be disengaged. Meanwhile, if the array were instead extended along the $z$ axis, the polarization of the scattered fields seen by the atoms would be identical to the incident field polarization, and no population would be pumped into state $|1\rangle$.

The system dynamics may be described by a master equation for the atomic density matrix $\rho$, given by $\dot{\rho} = -i[H,\rho] + \mathcal{L}(\rho)$, derived by formally integrating out the photonic modes \cite{WelschDipoleDipole,chang2012cavity,AnaGreenFunctionPaper}. The Hamiltonian $H = H_A + H_{\inter}$, where $H_A$ describes free atomic evolution and the external driving ($\hbar=1$):

\begin{equation}
H_A = -\sum_j\sum_g\sum_e \left(\Delta\sigma_{ee} + \Omega_{eg}\sigma_{eg}^j + \Omega^*_{eg}\sigma_{ge}^j\right).
\end{equation}

Here, $\Delta = \omega_L - \omega$ is the detuning of the incident field (of frequency $\omega_L$) from the resonant frequency $\omega$ of the ground to excited manifold transition. The atomic operator for the $j^{th}$ atom -- at position $\mathbf{r}_j$ -- is $\sigma_{\mu\nu}^j = |\mu_j\rangle\langle \nu_j|$, for energy eigenstates $|\mu\rangle$,$|\nu\rangle$. $\Omega_{eg}$ denotes the incident field Rabi frequency on the transition $|g\rangle\leftrightarrow|e\rangle$, and is given by $\Omega_{eg} = C_e^g (\hat{\bm{\epsilon}}_{-(eg)}\cdot\hat{\bm{\epsilon}}_L)d_{GE}E_0$. Here, $E_0$ is the incident field amplitude (equal for all atoms in our calculations), and $d_{GE}$ is the reduced dipole matrix element of the ground-excited manifold transition. $C_e^g = \langle J_g, m_g|J_e,m_e; 1, m_g-m_e\rangle$ denotes the Clebsch-Gordan coefficient of the transition connecting states $|g\rangle$ and $|e\rangle$ [Fig. \ref{SystemSetup}(b)], and the spherical basis vectors are written $\hat{\bm{\epsilon}}_{(eg)}$, where the subscript $(eg)$ denotes $m_{e} - m_{g}$, i.e. the difference in the corresponding magnetic quantum numbers. The circularly polarized components and spherical basis vectors are related by $\hat{\epsilon}_{\pm 1} = \hat{\sigma}_{\pm}$, and $\hat{\epsilon}_0=\hat{z}$.

The interaction Hamiltonian describes the photon-mediated exchange of excitations between atoms:

\begin{eqnarray}\label{FinalHeffInt}
H_{\inter} &=& -\frac{3\pi\Gamma}{k}\sum_{j,l}\sum_{g,g^{\prime}}\sum_{e,e^{\prime}}\left(\hat{\bm{\epsilon}}_{(e^{\prime}g^{\prime})}^*\cdot \mbox{Re}\:G^{jl}\cdot \hat{\bm{\epsilon}}_{(eg)}\right)\\ \nonumber
&&\qquad\qquad\qquad\qquad\times\: C_{e^{\prime}}^{g^{\prime}}\: C_{e}^{g}\:\sigma_{e^{\prime}g^{\prime}}^{j}\sigma_{ge}^{l}.
\end{eqnarray}

\vspace{5mm}

Here, $\Gamma = (d_{GE}C^{2}_6)^2k^3/3\pi\hbar\epsilon_0$ denotes the spontaneous decay rate of the transition $|6\rangle\rightarrow|2\rangle$. The electromagnetic Green tensor $G^{jl} = G(\mathbf{r}_{j},\mathbf{r}_l,\omega)$ physically describes the field at position $\mathbf{r}_j$ due to a classical oscillating dipole of frequency $\omega$ at position $\mathbf{r}_l$, and is given by \cite{novotny2006principles}

\begin{equation}\label{GreenTensor}
G_{\alpha\beta}(\mathbf{r}_j,\mathbf{r}_l,\omega) = \frac{1}{4\pi k^2}\left(k^2\delta_{\alpha\beta} + \partial_{\alpha}\partial_{\beta}\right)\frac{e^{ikr_{jl}}}{r_{jl}},
\end{equation}

\noindent where $\alpha,\beta = \{x,y,z\}$, $k = \omega/c$, and $r_{jl} = |\mathbf{r}_j - \mathbf{r}_l|$. The dissipative term in the master equation, which encodes collective spontaneous emission, is

\begin{eqnarray}\label{FinalLindblad}
\mathcal{L}(\rho) &=& -\frac{3\pi\Gamma}{k}\sum_{j,l}\sum_{g,g^{\prime}}\sum_{e,e^{\prime}}\left(\hat{\bm{\epsilon}}_{(e^{\prime}g^{\prime})}^*\cdot \mbox{Im}\:G^{jl}\cdot \hat{\bm{\epsilon}}_{(eg)}\right) \\
&\times& C_{e^{\prime}}^{g^{\prime}}\:C_{e}^{g}\left\{\sigma_{e^{\prime}g^{\prime}}^{j}\sigma_{ge}^{l}\rho + \rho\sigma_{e^{\prime}g^{\prime}}^{j}\sigma_{ge}^{l} - 2\sigma_{ge}^{l}\rho\sigma_{e^{\prime}g^{\prime}}^{j}\right\}.\nonumber
\end{eqnarray}

\vspace{5mm}

To demonstrate the presence of strong DDIs, we compute the steady-state population distribution in the ground manifold, employing two calculational methods. The first is a fully quantum approach, yielding the steady-state density matrix $\rho_{ss}$. Since the Hilbert space dimension grows with atom number $N$ as $6^N$, it is infeasible to compute directly with density matrices, and we instead use the quantum Monte Carlo wavefunction (QMCW) method, computing the evolution of the atomic wavevector $|\psi\rangle$ in individual quantum trajectories, and subsequently averaging over many trajectories to obtain an approximation of $\rho_{ss}$ (see Appendix~\ref{apa}).

\begin{figure}
\centering
\includegraphics[width=0.43\textwidth]{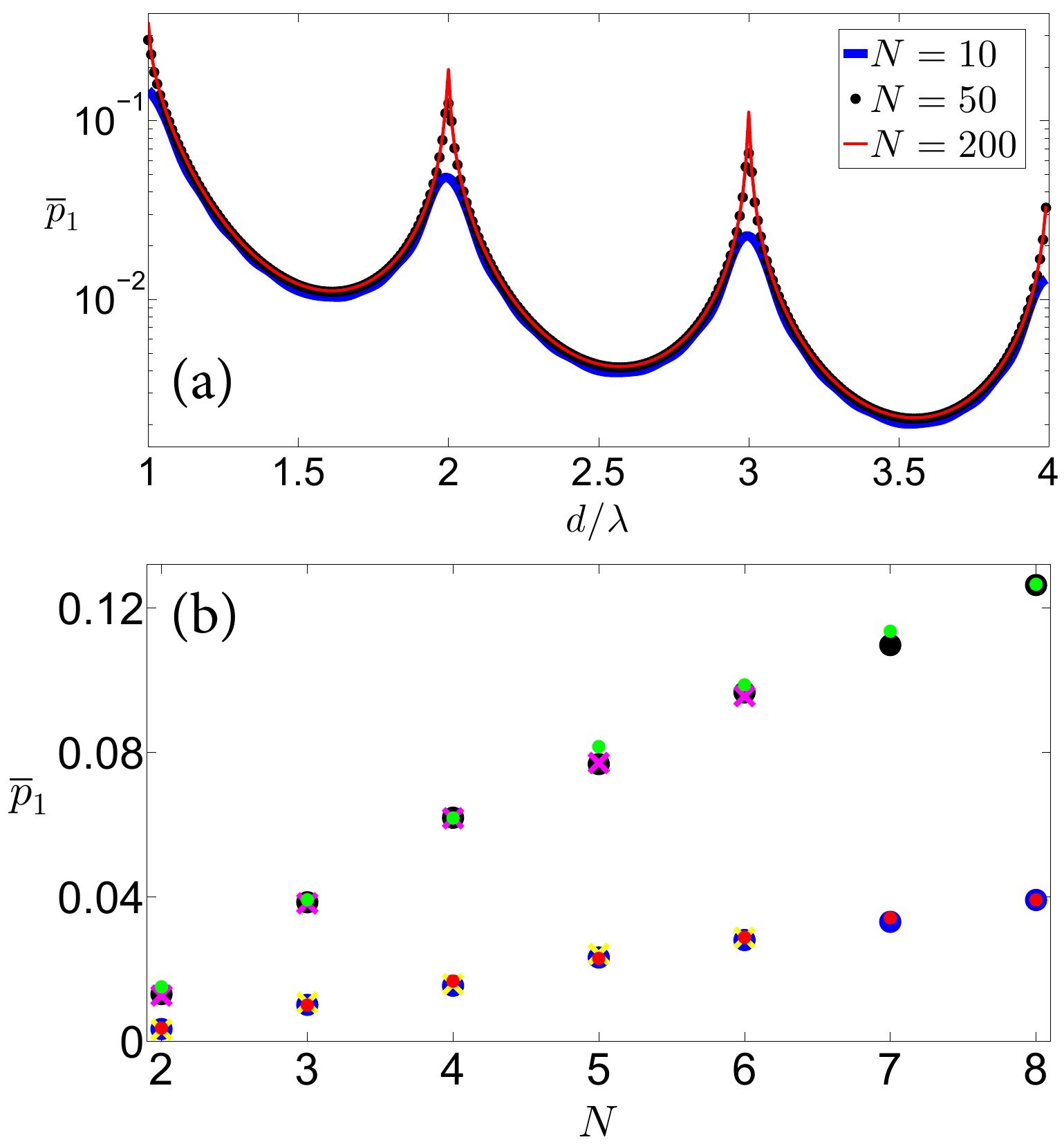}
\caption{(a) Mean steady-state population $\overline{p}_1$ vs. interatomic separation $d$, computed using the MF method, for chains of different atom number $N$ (see legend). (b) $\overline{p}_1$ vs. $N$ for $d=2\lambda$ and $d=\lambda$, computed using the QMCW method with the Hilbert space truncated at a single excitation (blue circles $d=2\lambda$; black circles $d=\lambda$), truncated at two excitations (yellow crosses $d=2\lambda$; magenta crosses $d=\lambda$), and the MF method (red circles $d=2\lambda$; green circles $d=\lambda$). Laser parameters: $\Omega/\Gamma = 0.01$, $\Delta/\Gamma=0$.}
\label{p1_FuncDist_n10n50n200}
\end{figure}

We will be interested in the case of weak driving: for all results presented we take the Rabi frequency on the transition between states $|2\rangle$ and $|6\rangle$ to be $\Omega_{26} \equiv \Omega = 0.01\Gamma$. The small population in the excited manifold then allows truncation of the Hilbert space at one or two total excitations. Even within this truncated regime, however, the restricted subspace grows exponentially in the number of atoms, since the ground manifold alone has dimension $2^N$.

To circumvent limitations on $N$ arising from the large Hilbert space, the second method we employ is a mean field (MF) approach. The time evolution of the expectation value of any atomic observable $\langle\sigma^i_{\mu\nu}\rangle$ in general depends on two-atom correlation functions of the form $\langle\sigma^i_{\mu^{\prime}\nu^{\prime}}\sigma^j_{\mu^{\prime\prime}\nu^{\prime\prime}}\rangle$. We assume such correlation functions may be factorized into products of single-atom expectation values, $\langle\sigma^i_{\mu^{\prime}\nu^{\prime}}\sigma^j_{\mu^{\prime\prime}\nu^{\prime\prime}}\rangle \approx \langle\sigma^i_{\mu^{\prime}\nu^{\prime}}\rangle\langle\sigma^j_{\mu^{\prime\prime}\nu^{\prime\prime}}\rangle$. The resulting equations of motion may be derived from an effective Hamiltonian $H = H_A+H_{MF}$, where

\begin{equation}
H_{\mbox{\scriptsize MF}} = -\sum_j \sum_{g,e} \left(\mathcal{R}_{eg}^j\sigma_{eg}^j  + \mathcal{R}_{eg}^{j\:*}\sigma_{ge}^j\right).
\end{equation}

Here we define an effective Rabi frequency on the $|e\rangle\leftrightarrow|g\rangle$ transition of the $j^{th}$ atom, given by

\begin{equation}\label{EffRabiFreq}
\mathcal{R}_{eg}^j = C_{e}^{g}\:\hat{\bm{\epsilon}}_{(eg)}^*\cdot\left[\frac{3\pi\Gamma}{k}\sum_{\substack{l\neq j\\g^{\prime},e^{\prime}}}\: C_{e^{\prime}}^{g^{\prime}}\:G^{jl}\cdot \hat{\bm{\epsilon}}_{(e^{\prime}g^{\prime})}\: \langle \sigma_{g^{\prime}e^{\prime}}^{l} \rangle\right].
\end{equation}

Within the MF approach, all other atoms $l\neq j$ are self-consistently treated as classical coherent external driving sources for the $j^{th}$ atom. Single atom emission processes are described by the Lindblad operator

\begin{equation}
\mathcal{L}_{\ind}(\rho) = -\frac{\Gamma}{2}\sum_{j,g,e}(C^g_e)^2\left(\sigma_{ee}^j\rho + \rho\sigma_{ee}^j - 2\sigma_{ge}^j\rho\sigma_{eg}^j\right).
\end{equation}

\section{RESULTS AND INTERPRETATION}

We first apply the MF approach to investigate the mean steady-state population $\overline{p}_1$ of ground state $|1\rangle$, averaged over all $N$ atoms, as a function of the lattice spacing $d$. For an incident field resonant with the transition from the ground to excited manifold ($\Delta = 0$), Fig. \ref{p1_FuncDist_n10n50n200}(a) shows $\overline{p}_1$ vs. $d$ for chains containing $N=10$, $N=50$ and $N=200$ atoms. As anticipated above, at integer multiples of the resonant wavelength $\lambda$, where the scattered fields interfere constructively, sharp maxima are observed in $\overline{p}_1$. Specifically, when $N=200$ and $d=2\lambda$, $\overline{p}_1$ accounts for $\approx$ 20\% of the total atomic population, clearly signalling a breakdown of pumping in the absence of DDIs.

The validity of the MF approximation is justified by comparison with the results of the QMCW approach. Figure \ref{p1_FuncDist_n10n50n200}(b) shows the predictions of both methods for $\overline{p}_1$ vs. $N$, for spacings of $2\lambda$ and $\lambda$. The QMCW calculations, with Hilbert spaces truncated to both one and two excitations, agree very well with mean field to the largest atom numbers $N$ that can be feasibly calculated. As discussed in Appendix~\ref{apb}, further analysis shows that the MF approximation works well down to $d\sim \lambda/2$. This is consistent with previous studies of 1D arrays, which have shown that long-lived (subradiant) collective excitations are supported when $d<\lambda/2$, resulting from strong atom-atom interactions \cite{SelectiveSubradiance}. Intuitively, one would expect the MF approximation to break down in this regime, due to the presence of quantum correlations.

\begin{figure}
\centering
\includegraphics[width=0.48\textwidth]{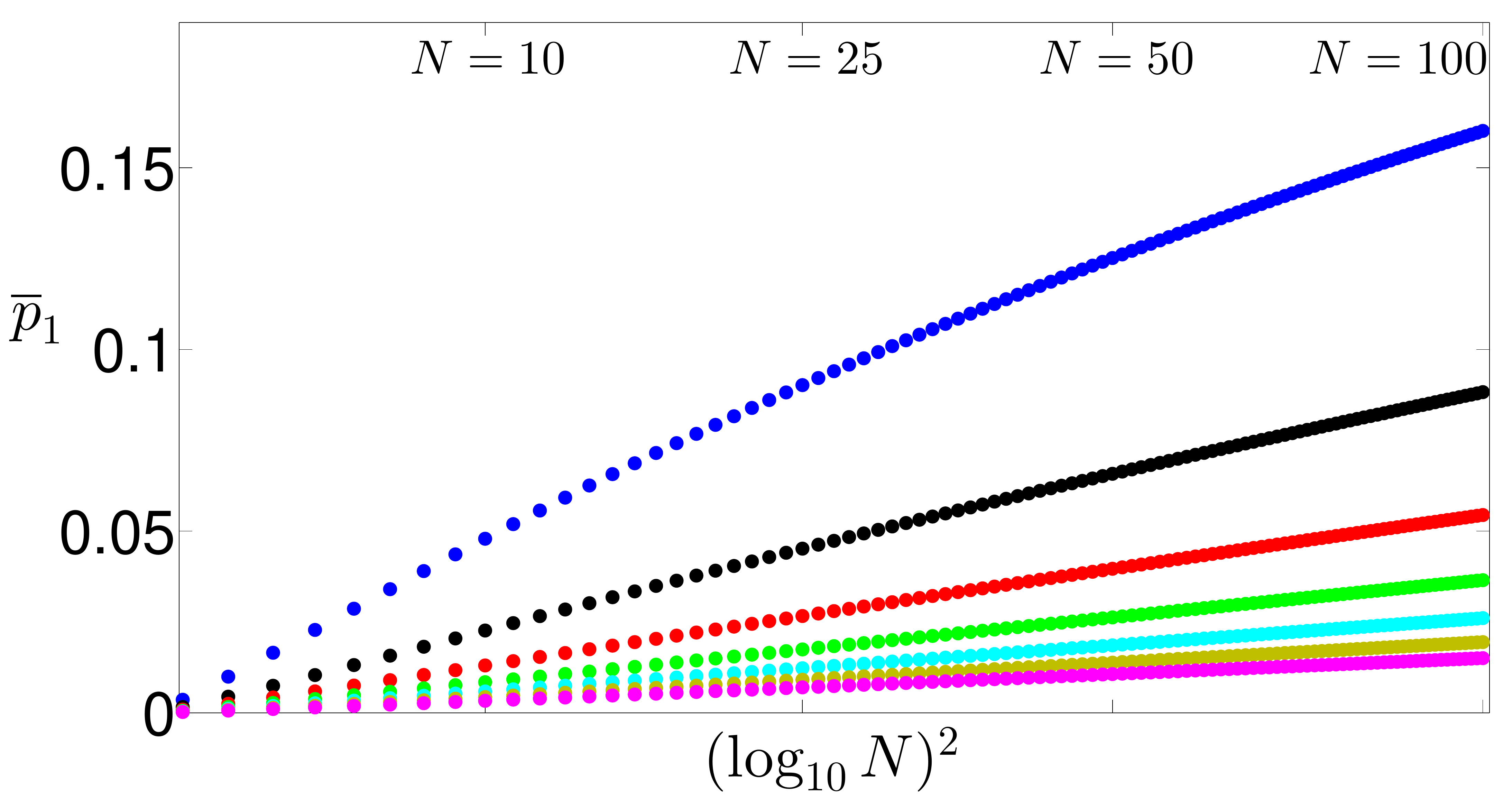}
\caption{Scaling of $\overline{p}_1$ vs. $(\log_{10}N)^2$, for up to $N=100$ atoms, and for lattice spacings $d/\lambda = 2,3,...,8$, from top (blue) to bottom (magenta). The $x$-axis tick marks indicate the cases $N = \{10,25,50,100\}$, as shown in the upper part of the figure. Laser parameters: $\Omega/\Gamma = 0.01$, $\Delta/\Gamma=0$.}
\label{p1_func_logsqn_linear_Ver4}
\end{figure}

Having verified the validity of the MF model, we now study how $\overline{p}_1$ scales with $N$, when the spacing $d$ is an integer multiple of $\lambda$ such that constructive interference is prominent. Figure \ref{p1_func_logsqn_linear_Ver4} shows how $\overline{p}_1$ grows as $(\lambda/d)^2\log^2 N$. To explain this, we first note that for a single atom driven with both $\sigma_+$ and $\sigma_-$ external fields, of respective amplitudes $E_+$ and $E_-$, the steady-state ratio of the ground state populations is $p_1/p_2 = |E_-/E_+|^2$. For our problem we then expect that $\overline{p}_1 = |E_-^{\mbox{\scriptsize sc}}/(E_+^{\mbox{\scriptsize inc}} + E_+^{\mbox{\scriptsize sc}})|^2\overline{p}_2$, where the superscripts $inc$ and $sc$ denote incident and scattered fields. When the scattering is weak, we further expect that $\overline{p}_1 \approx |E_-^{\mbox{\scriptsize sc}}/E_+^{\mbox{\scriptsize inc}}|^2$, since $\overline{p}_2\approx 1$. Moreover, the atoms radiate predominantly on the $|2\rangle - |6\rangle$ transition, which is driven by the incident field. As explained above, this dipole produces a $y$-polarized far field along the array axis, with amplitude $\sim e^{ikr}/r$ a distance $r$ away [Eq. (3)]. When $kr$ is an integer multiple of $2\pi$ (i.e. when $d$ is an integer multiple of $\lambda$), these fields add constructively. In particular, the scattered field experienced by a typical atom -- originating from all other atoms -- is $E_{-}^{sc}\sim (\lambda/d) \sum_{j=1}^{N} (1/j) \sim (\lambda/d)\log N$, which explains the observed scaling behavior. This simple argument breaks down when $N$ is sufficiently large that the scattered and incident field amplitudes are comparable, and depletion of population from state $|2\rangle$ cannot be neglected. For spacings $d$ different from an integer multiple of $\lambda$, $\overline{p}_1$ is independent of $N$ for $N\gg1$, as discussed in Appendix~\ref{apb}.

\begin{figure}
\centering
\includegraphics[width=0.5\textwidth]{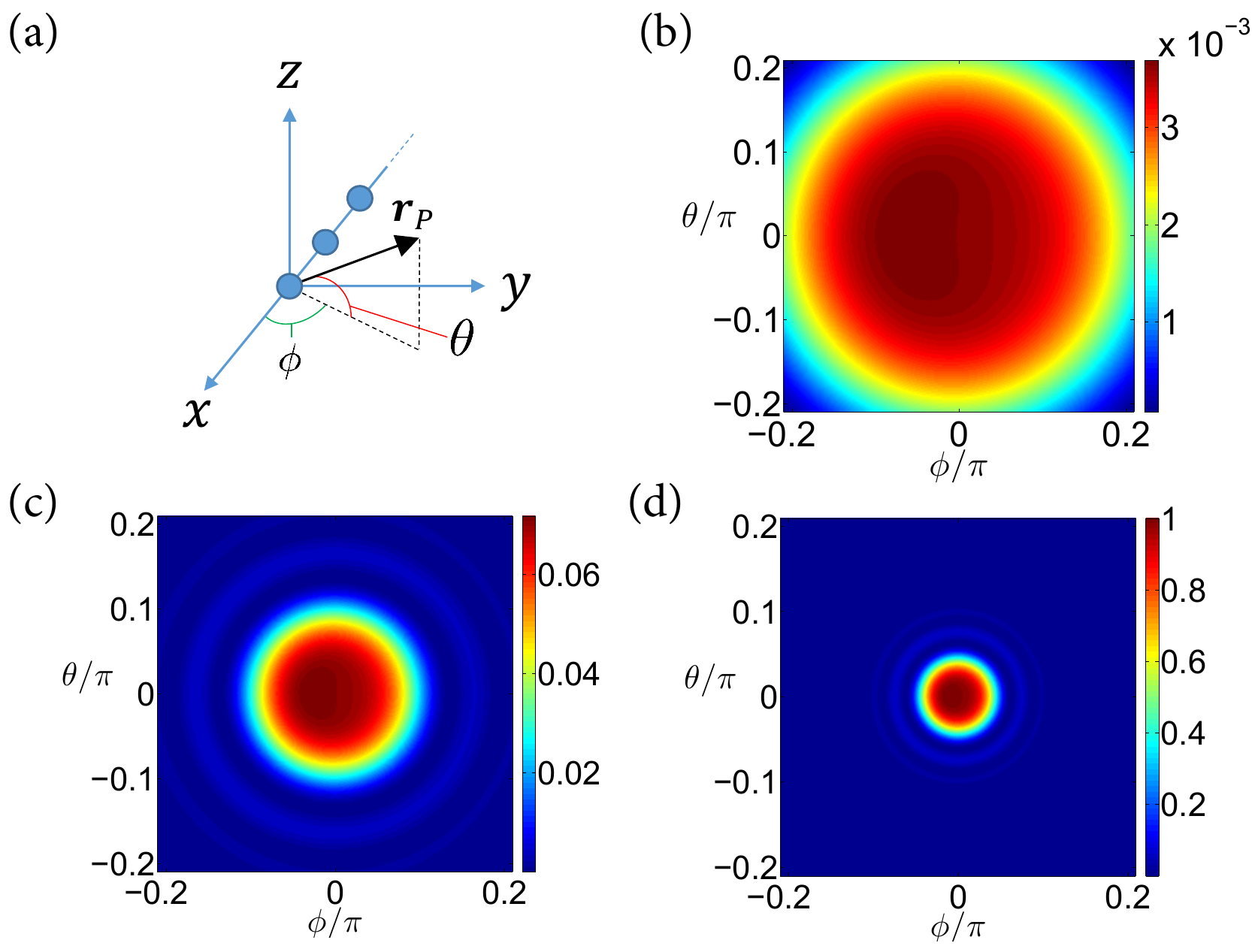}
\caption{The total far-field intensity computed at position $\mathbf{r}_P$ relative to the atom with the largest $x$ coordinate, as a function of the angles $\theta$ and $\phi$, defined in (a). The intensity is plotted for (b) $N=10$, (c) $N=50$, and (d) $N=200$, and has been normalized to the maximum intensity of the case $N=200$. Other parameters: $d/\lambda = 2$, $\Omega/\Gamma = 0.01$, and $\Delta/\Gamma=0$.}
\label{Intensity_50AtomChain}
\end{figure}

In sufficiently large arrays with $d=m\lambda$ for integer $m$, the scattered field amplitude at the atomic positions can exceed the incident field amplitude. This does not violate energy conservation, since the solid angle within which constructive interference occurs becomes increasingly small with larger $N$, ensuring the emitted intensity does not exceed the input intensity. This is illustrated in Figs. \ref{Intensity_50AtomChain}(b)-(d), which show the emitted field intensity profile in the far-field region for systems with $N = (50,100,200)$, normalized to the maximum intensity of the case $N=200$. Ultimately, therefore, one expects that as $N\rightarrow\infty$, the steady-state population should become equally distributed across the ground states $|1\rangle$ and $|2\rangle$, i.e. $\overline{p}_1 = \overline{p}_2$. While formally true, this saturation behavior occurs only for unrealistically large systems, where $N\sim 10^9$ -- see Appendix~\ref{apb}. Finally, we note that Figs.~\ref{Intensity_50AtomChain}(b)-(d) display a small asymmetry about the array axis: we attribute this to the relative phases between the field components radiated by the different atomic coherences.

\section{DISORDER IN THE ATOMIC POSITIONS}

The results presented above assume that the atoms form a perfectly ordered array, with a fixed distance $d$ between neighboring atoms. However, within each tweezer there is an uncertainty in the atomic position. In the following, we will consider the effect of such random position disorder on the previous results. Intuitively, one expects that the role of disorder is to degrade the strong constructive interference of the emitted fields, resulting in a reduction of the population $\overline{p}_1$ acquired in the ground state $|1\rangle$.

We will consider a simple model, where the position disorder is taken to be solely along the axis of the array. Specifically, we define $d_{i,i+1}$ to be the separation between the $i^{th}$ atom and its nearest neighbor, where $d_{i,i+1} = (2+\xi_{i,i+1})\lambda$. Here, $\xi_{i,i+1}$ is a small disorder parameter drawn randomly from the interval $(0,\epsilon)$, where $\epsilon$ represents the disorder `strength'. Given a set of such $\xi_{i,i+1}$ for all neighbouring pairs of atoms, we compute the population $\overline{p}_1$ as explained in the previous section. We repeat the computation for 500 different sets of $\xi_{i,i+1}$, each yielding a characteristic value for $\overline{p}_1$. Finally, we compute the mean of the resulting 500 values of $\overline{p}_1$ to obtain the disorder-averaged population, which we denote $\langle\overline{p}_1\rangle$.

\begin{figure}
\centering
\includegraphics[width=0.49\textwidth]{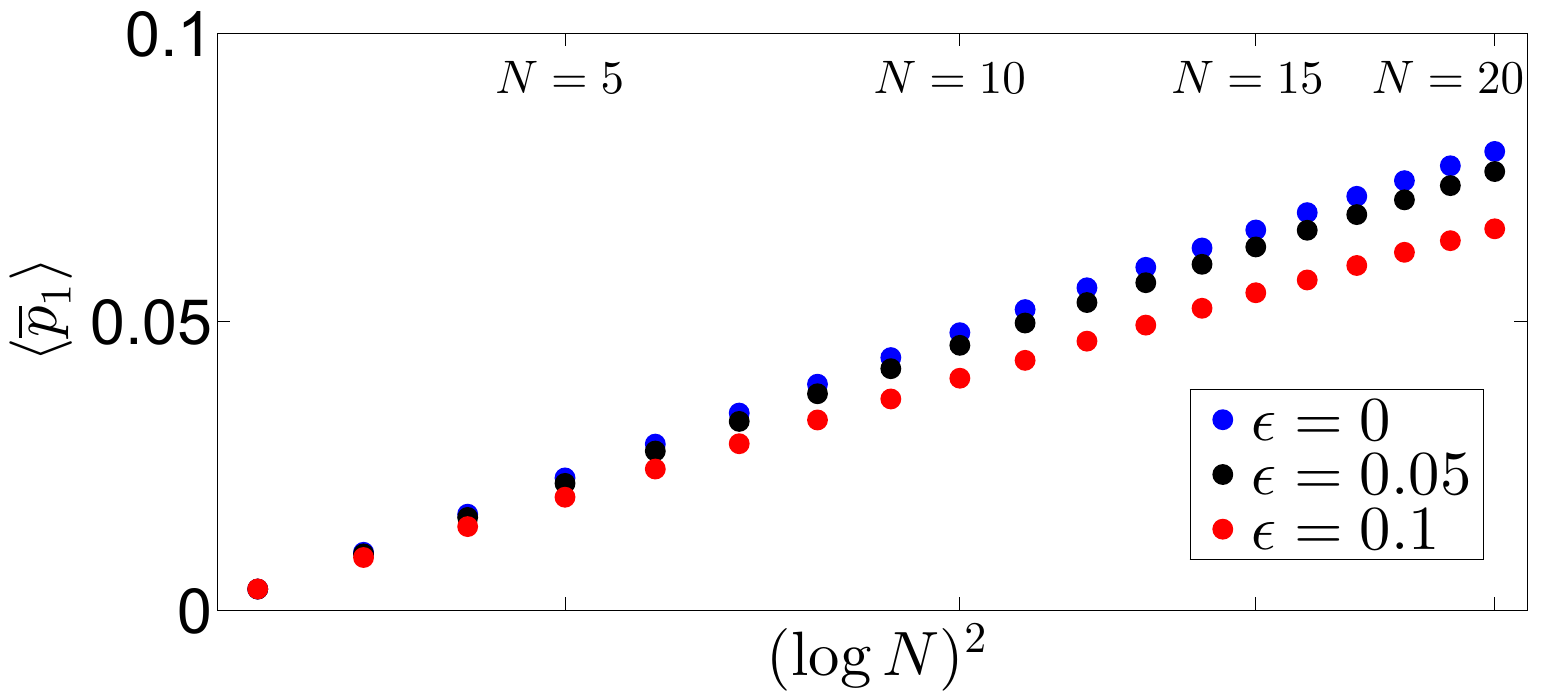}
\caption{Population $\langle\overline{p}_1\rangle$ as a function of the number of atoms $N$, averaged over atomic positions for different disorder strengths $\epsilon$.}
\label{DisorderFigure}
\end{figure}

Figure \ref{DisorderFigure} shows $\langle\overline{p}_1\rangle$ as a function of the number of atoms $N$ for different $\epsilon$. As anticipated, the population transferred to state $|1\rangle$ is reduced, however for realistic parameters the effect is relatively small, and the qualitative scaling of $\langle\overline{p}_1\rangle$ with $N$ described in the previous section is unaffected.

\section{CONCLUSION AND OUTLOOK}

Our results demonstrate that strong dipole-dipole interactions due to multiple scattering exhibit unique signatures in ordered arrays of multilevel atoms, and moreover in a regime accessible to current experiments, $d>\lambda$. We expect our findings to be of general interest to experiments where strong collective effects in atomic ensembles are desirable (e.g. in engineering correlated atomic states or output fields), and also where they are deleterious (e.g. in achieving efficient optical pumping). Additionally, our work could be extended to study residual light scattering in arrays of trapped ions \cite{BlattIonTrapQIP,Monroe53qubit} and neutral atoms \cite{Lukin51Qubit}, which constitute particularly promising routes towards high-fidelity quantum simulation, as well as to quantify errors in optical lattice clocks resulting from DDIs \cite{JunYeOpticalClocksReview}. In the context of atomic arrays, in light of our results it would be interesting to revisit phenomena such as perfect reflection from 2D lattices, and subradiance in 1D chains, to understand how multilevel structure may affect results previously obtained for two-level atoms. Finally, we anticipate the presence of strong quantum correlations in the ground manifold for sub-wavelength lattice constants, whose nature warrants further study.\\

\textbf{Acknowledgments --} The authors thank HJ Kimble for stimulating discussions. EM and LCK acknowledge support from the National Research Foundation, Prime Minister's Office, Singapore and the Ministry of Education, Singapore under the Research Centres of Excellence programme. AAG was supported by an IQIM postdoctoral fellowship and the Global Marie Curie Fellowship LANTERN. CAR and YL acknowledge support from the Office of Naval Research, AFOSR MURI under Grant No. FA9550-16-1-0323, and the NSF under grant number PHYS 1734006. DEC acknowledges support from the ERC Starting Grant FOQAL, MINECO Plan Nacional Grant CANS, and MINECO Severo Ochoa Grant SEV-2015-0522, CERCA Programme / Generalitat de Catalunya, and Fundacio Privada Cellex.

\appendix
\section{QUANTUM MONTE CARLO METHOD}\label{apa}

In the main text we noted that the dimension of the full Hilbert space of our system is $6^N$, where $N$ is the number of atoms. The ground state manifold alone has a dimension that scales exponentially in the number of atoms, $2^N$, while the first excited manifold has dimension $4N\times 2^{N-1}$. Clearly, even within the weak driving limit we have considered, where we work only in the reduced subspace of zero and one excitations, the Hilbert space becomes very large for modest numbers of atoms $N$. It is then infeasible to compute using density matrices, and we instead turn to the quantum Monte Carlo wavefunction (QMCW) approach.

Conventional implementations of the QMCW algorithm \cite{Dum_QMCW,MolmerQMCW,DaleyQuantumJumps} are not optimal for our problem, since for weak driving the time to reach steady state $\tau_{ss}$ is orders of magnitude larger than the time scale of decay, which is $\sim 1/\Gamma$. However, quantum jumps are also very infrequent, a feature which we may exploit by taking large time steps using the time evolution operator $U(\Delta t) = \exp(-iH_{\eff}\Delta t)$. Here, the effective Hamiltonian $H_{\eff} = H - i\sum_l \gamma_l J_l^{\dagger}J_l$, where $H$ is given by Eq. (1) of the main text, and $J_l$ denotes a set of operators describing the possible quantum jumps that can occur, with corresponding rates $\gamma_l$. In order to obtain the operators $J_l$ and rates $\gamma_l$, we diagonalize the matrix of coefficients appearing in the dissipative part of the master equation (Eq. (4) in the main text), whose entries are formed from the imaginary part of the Green tensor $\hat{\epsilon}_{q^{\prime}}^*\cdot\mbox{Im}\:G(r_j,r_l)\cdot\hat{\epsilon}_q$ and the appropriate Clebsch-Gordan coefficients.

We first generate a random number $x$ from the uniform distribution $x \in (0,1)$, as per the conventional QMCW algorithm. We then evolve the state vector of the system $|\psi(t)\rangle$ from an initial time $t$ to a later time $|\psi(t+\Delta t)\rangle$ using a combination of matrix exponentials as follows. First, a large step $t_l = 1000/\Gamma$ is applied, and the loss of probability $\Delta p$ due to the non-Hermitian evolution is computed, given by $\Delta p = 1 - |\langle \psi(t+t_l)|\psi(t+t_l)\rangle|$. If we find $\Delta p<x$, we simply renormalize the wavefunction, and then apply the large time step evolution operator again to this updated wavefunction.

On the contrary, if after the first time step $t_l$ we find $\Delta p>x$, we return to the initial state $|\psi(t)\rangle$, and evolve it using a medium time step $t_m = 25/\Gamma$. We again compute $\Delta p$, and if $\Delta p<x$, we renormalize the resulting wavefunction and evolve again for a time $t_m$. If $\Delta p>x$, however, we return to the initial state $|\psi(t)\rangle$ and evolve it by the small time step $t_s = 1/\Gamma$, until we finally find the time at which $\Delta p$ becomes greater than $x$. We then collapse the wavefunction with one of the jump operators $J_l$.

In order to determine which jump takes place, we calculate the relative probability $p_l$ of each of them occurring,

\begin{equation}
p_l = \frac{\langle \psi(\tau)|J_l^{\dagger}J_l|\psi(\tau)\rangle}{\sum_k\langle \psi(\tau)|J_k^{\dagger}J_k|\psi(\tau)\rangle},
\end{equation}

where $\tau$ denotes the time at which the jump happens. Based on the values of the $p_l$, we may assign each jump operator a corresponding proportion of the interval $(0,1)$, and then generate a second random number $y$ -- again drawn from the uniform distribution $(0,1)$ -- whose value is then used to determine which jump occurs. We repeat this procedure until the desired final time $t_f$ is reached, thus obtaining one possible trajectory for the system evolution. We then repeat the method $n_{\mbox{\scriptsize traj}}$ times, and obtain an approximation to the full density matrix via

\begin{equation}
\rho(t_f) \approx \frac{1}{n_{\mbox{\scriptsize traj}}}\sum_j|\psi_j(t_f)\rangle\langle\psi_j(t_f)|,
\end{equation}

where $j=1,...,n_{\mbox{\scriptsize traj}}$ labels the individual trajectories. For the results presented in the main text, $n_{\mbox{\scriptsize traj}} = 2400$.

\begin{figure}
\centering
\includegraphics[width=0.49\textwidth]{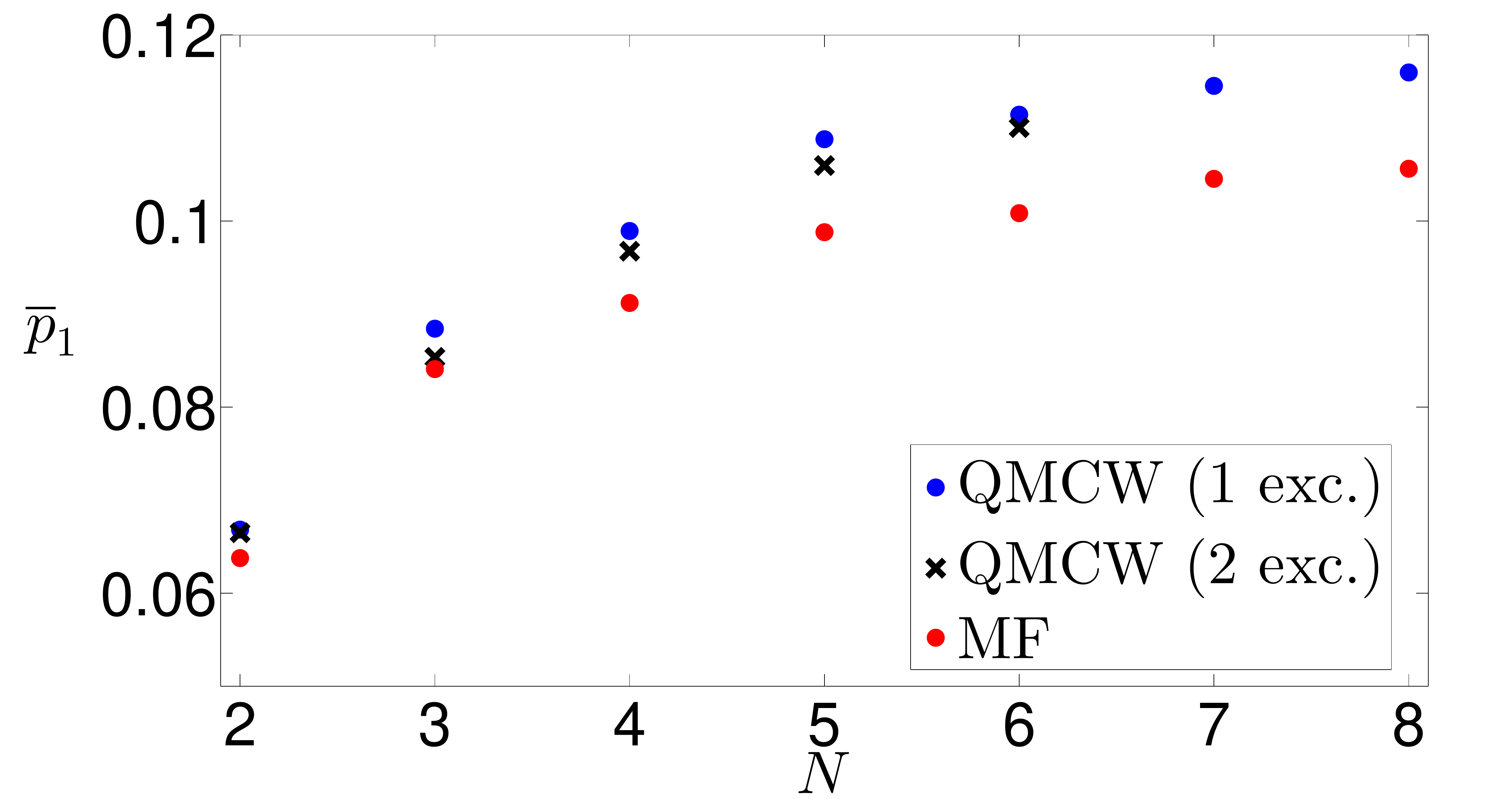}
\caption{Steady-state population $\overline{p}_1$ of state $|1\rangle$, averaged over all atoms for the case $d=\lambda/2$, comparing the predictions of the QMCW method (truncated at a single excitation up to $n=8$, and at two excitations up to $n=6$) and the mean field (MF) method.  Laser parameters: $\Omega/\Gamma = 0.01$, $\Delta/\Gamma=0$.}
\label{p1CompareQM_MF_2}
\end{figure}

Note that our choice of $t_s$ for the short time step places a limit on how finely we can resolve the time that a given quantum jump occurs: specifically, we may only resolve to within one atomic lifetime. This will introduce a certain amount of error, because: (a) it is possible that two jumps occur within one lifetime, or; (b) the relative probabilities of different jumps occurring may vary over a time scale of the order of $1/\Gamma$, and thus when we determine which jump to apply, we may make an error and apply the wrong one.

However, given the weak driving and the small probability of atoms being excited in our system, the time between successive jumps should intuitively be very large compared to the time required for the system to re-equilibrate after a jump; we have confirmed this by numerically by analyzing the time distribution of jumps. It then follows that errors of type (a) will be very small. It also follows that when a jump does occur, with high probability the system state immediately before the jump is the quasi-steady state, in which case the relative jump probabilities will be stable. Errors of type (b) will therefore also be very small.

\section{MEAN FIELD METHOD - SUPPLEMENTARY RESULTS}\label{apb}

\subsection{Mean field solution for $d = \lambda/2$}

In Fig. 2(b) of the main text we showed that the mean field and QMCW methods agree in their predictions for the population $\overline{p}_1$ for the cases $d = 2\lambda$ and $d=\lambda$. Figure \ref{p1CompareQM_MF_2} shows that the mean field model remains approximately consistent with the QMCW method even for the case $d=\lambda/2$. As noted in the main text, in 1D subradiant effects are only prominent for separations $d<\lambda/2$, where one therefore expects strong quantum correlations to occur \cite{SelectiveSubradiance}. For $d > \lambda/2$, the corresponding absence of strong correlations implies that a mean field approach should be a good approximation to the full quantum master equation.

\subsection{Atoms separated by non-integer multiples of a wavelength}
\begin{figure}
\centering
\includegraphics[width=0.49\textwidth]{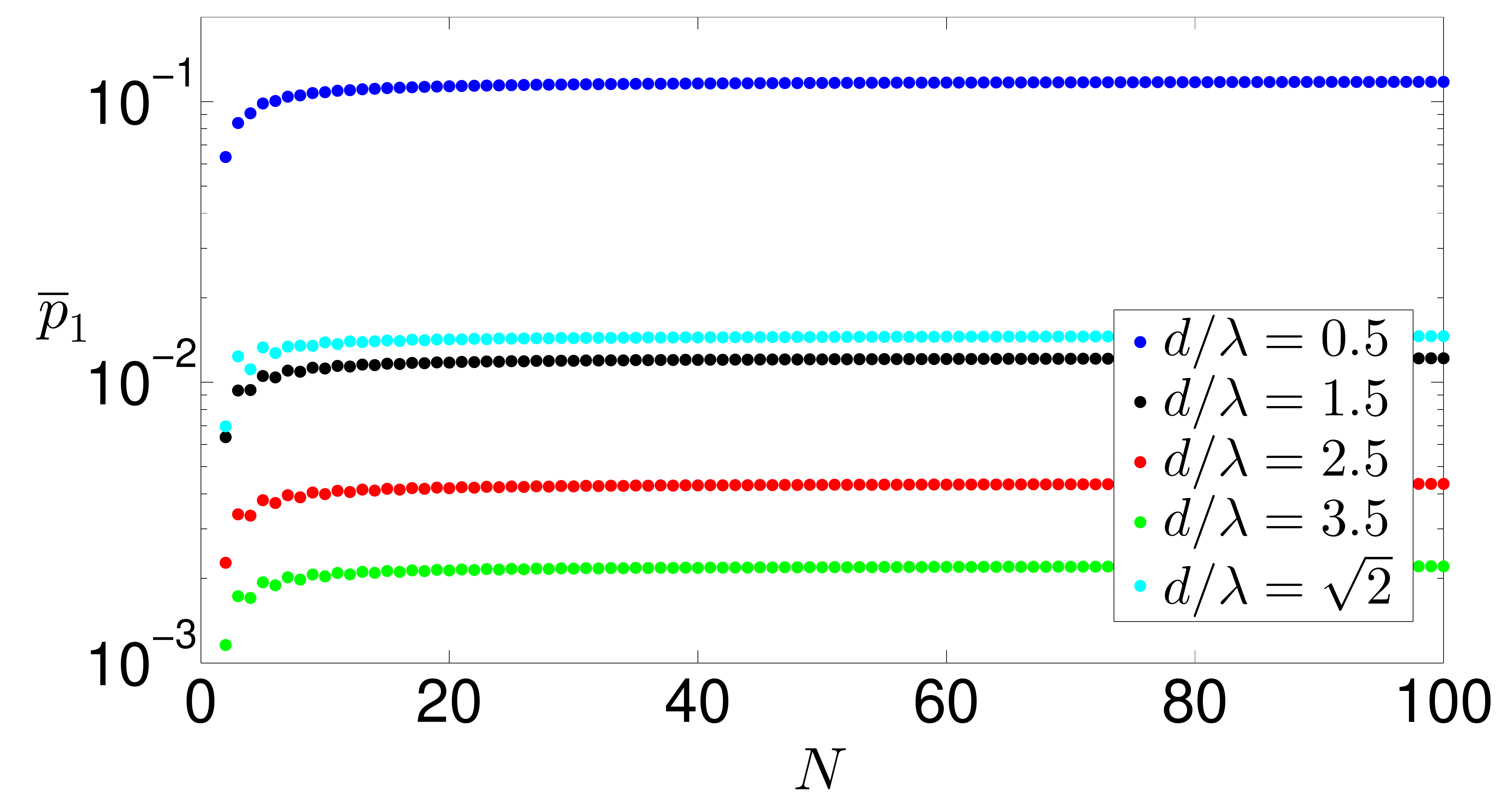}
\caption{Scaling of the population $\overline{p}_1$ vs. $N$, for $N\in(1,100)$, and for different interatomic separations $d$. Laser parameters: $\Omega/\Gamma = 0.01$, $\Delta/\Gamma=0$.}
\label{p1Func_n_d_halflambda}
\end{figure}

Provided that the lattice spacing is not close to an integer multiple of a resonant wavelength $\lambda$, from Fig. 2 of the main text one sees that the population $\overline{p}_1$ is largely independent of the number of atoms. This is shown explicitly in Fig. \ref{p1Func_n_d_halflambda}, where $\overline{p}_1$ is plotted as a function of atom number $N$ for systems whose lattice constant satisfies $d=(m+1/2)\lambda$ for integer $m$, as well as for a system with $d = \sqrt{2}\lambda$ where the spacing is an irrational multiple of the wavelength. In both cases, the population $\overline{p}_1$ is seen to converge to a constant value, which may be explained by arguments similar to those given in the main text.

In particular, for the case $d=(m+1/2)\lambda$ the $\sigma_-$-polarized scattered field is proportional to an alternating harmonic series

\begin{equation}
E_{-}^{sc}\sim \frac{\lambda}{d} \sum_{j=1}^{n}\frac{(-1)^j}{j}.
\end{equation}

The sum converges to a value of $\log 2$, thereby implying that for sufficiently large $N$ -- seen from Fig. \ref{p1Func_n_d_halflambda} to be around $N\gtrsim 20$ -- the population $\overline{p}_1$ is determined only by the lattice constant $d$, i.e. $\overline{p}_1 \sim (\lambda/d)^2$. Similar behavior is observed for the irrational lattice constant $d=\sqrt{2}\lambda$: in this case, the field $E_{-}^{sc}$ is proportional to a random harmonic series, where the numerator of each term is simply a random complex number. This series also converges, so that again $\overline{p}_1 \sim (\lambda/d)^2$ for sufficiently large $N$.

\subsection{Very large $N$ limit}
\begin{figure}
\centering
\includegraphics[width=0.49\textwidth]{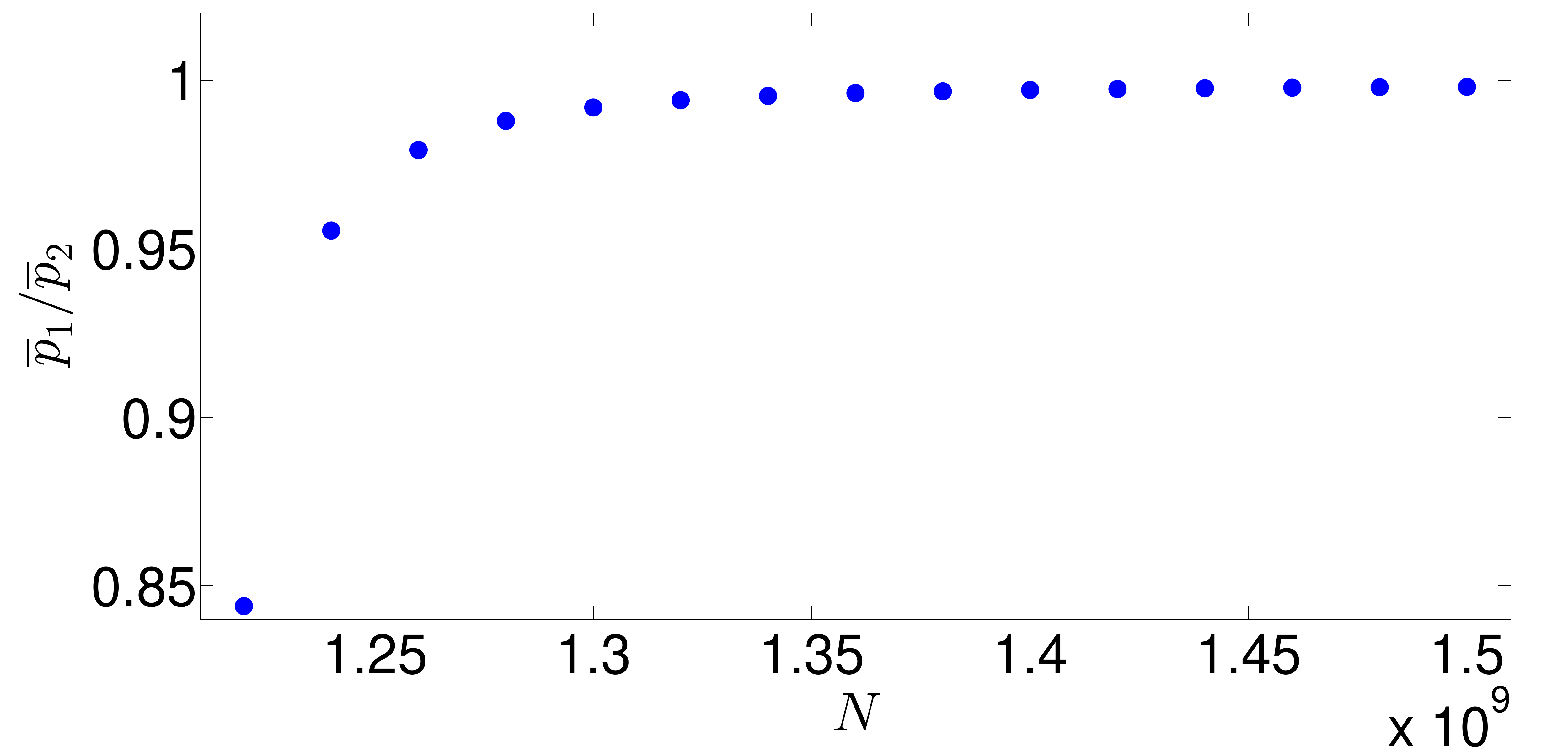}
\caption{Ratio $\overline{p}_1/\overline{p}_2$ as a function of $N$ in the limit where the system size is very large, and the mean field equations reduce to a single-atom problem as described in the text. Here, the lattice spacing is $d=2\lambda$, and the laser parameters are $\Omega/\Gamma = 0.01$, $\Delta/\Gamma=0$.}
\label{MeanField_Ratio_p1p2_verylarge_n}
\end{figure}

In the regime where the number of atoms becomes very large, it is possible to simplify the mean field equations significantly by assuming that all atoms in the system behave identically. In particular, the mean value of a given coherence $\langle \sigma_{ge}\rangle$ is equal for all atoms, and the full $N$-body problem is reduced to solving for the variables of just a single atom; the $N$ dependence in this case then enters via the sum over Green tensor elements in Eq. (6) of the main text. Figure \ref{MeanField_Ratio_p1p2_verylarge_n} shows the steady-state ratio $\overline{p}_1/\overline{p}_2$ as a function of $N$ in this regime, which tends to unity for very large arrays. The number of atoms required to reach this limit, however, is seen to be unrealistically large.

\bibliographystyle{apsrev4-1}
\bibliography{references}

\end{document}